\documentclass[prl,twocolumn,showpacs,superscriptaddress]{revtex4}

\usepackage{graphicx}
\usepackage{dcolumn}
\usepackage{bm}

\usepackage{color}
\newcommand{\red}{\color{red}}

\begin{document}


\title{Spin-polarized quasi 1D state with finite bandgap on the Bi/InSb(001) surface}

\author{J. Kishi}
\affiliation{Department of Physics, Graduate School of Science, Osaka Unviersity, Toyonaka 560-0043, Japan}
\author{Y. Ohtsubo}
\email{y_oh@fbs.osaka-u.ac.jp}
\affiliation{Graduate School of Frontier Biosciences, Osaka University, Suita 565-0871, Japan}
\affiliation{Department of Physics, Graduate School of Science, Osaka Unviersity, Toyonaka 560-0043, Japan}
\author{T. Nakamura}
\affiliation{Department of Physics, Graduate School of Science, Osaka Unviersity, Toyonaka 560-0043, Japan}
\author{K. Yaji}
\author{A. Harasawa}
\author{F. Komori}
\author{S. Shin}
\affiliation{Institute for Solid State Physics, The University of Tokyo, 5-1-5 Kashiwanoha, Kashiwa, Chiba 277-8581, Japan}
\author{J. E. Rault}
\author{P. Le F\`evre}
\author{F. Bertran}
\author{A. Taleb-Ibrahimi}
\affiliation{Synchrotron SOLEIL, Saint-Aubin-BP 48, F-91192 Gif sur Yvette, France}
\author{M. Nurmamat}
\affiliation{Graduate School of Science, Hiroshima University, 1-3-1 Kagamiyama, Higashi-Hiroshima 739-8526, Japan}
\author{H. Yamane}
\author{S. Ideta}
\author{K. Tanaka}
\affiliation{Institute for Molecular Science, Okazaki 444-8585, Japan}
\author{S. Kimura}
\email{kimura@fbs.osaka-u.ac.jp}
\affiliation{Graduate School of Frontier Biosciences, Osaka University, Suita 565-0871, Japan}
\affiliation{Department of Physics, Graduate School of Science, Osaka Unviersity, Toyonaka 560-0043, Japan}
\date{\today}

\begin{abstract}
One-dimensional (1D) electronic states were discovered on 1D surface atomic structure of Bi fabricated on semiconductor InSb(001) substrates by angle-resolved photoelectron spectroscopy (ARPES).
The 1D state showed steep, Dirac-cone-like dispersion along the 1D atomic structure with a finite direct bandgap opening as large as 150 meV.
Moreover, spin-resolved ARPES revealed the spin polarization of the 1D unoccupied states as well as that of the occupied states, the orientation of which inverted depending on the wave vector direction parallel to the 1D array on the surface.
These results reveal that a spin-polarized quasi-1D carrier was realized on the surface of 1D Bi with highly efficient backscattering suppression, showing promise for use in future spintronic and energy-saving devices.
\end{abstract}

\maketitle

For the past decade, electronic states showing Dirac cone (DC)-type dispersion with very steep dispersion and extremely high carrier mobility have been one of the topics of much interest in solid-state physics \cite{Geim07, Hasan10}.
The surface states of topological insulators (TI) \cite{Hasan10} are regarded as promising candidates for future spintronic technologies because of their characteristic helical spin polarization and suppression of backscattering (BS) \cite{Manchon15, Roushan09}.

The suppression of BS on topological surface states is due to its typical spin-polarized structure.
Since non-magnetic scatterers cannot cause any spin inversion in elastic scattering, direct BS is forbidden in spin-polarized Dirac states on TIs (Fig. 1(a)) \cite{Roushan09}.
Because of this BS suppression, high-speed and dissipation-free conduction via topological surface states is expected.
However, a two-dimensional (2D) state is not suitable to achieve a perfect dissipation-free conduction because of the scattering towards obliquely backward directions which results in a finite non-parallel spin component between the initial/final states \cite{Kim14}.
Such obliquely BS is absent for 1D system such as the 1D topological edge states on 2D TIs \cite{Konig07}.
However, the details of the electronic structure of such edge states, such as band dispersion and spin-orbital polarizations, is not clear yet because of experimental difficulty to access the 1D state localized in the very small area on the edge of 2D materials.

In parallel to the 2D TIs, the quasi-1D (Q1D) states in highly anisotropic materials have been reported recently, such as vicinal surfaces of Bi single crystal \cite{Wells09, Bianchi15} and Bi$_4$I$_4$ which is a 3D TI with a Q1D lattice \cite{Autes16}.
In both cases, the Q1D surface states showed DC-like very steep dispersion.
Moreover, clear spin polarization is observed normal to the wave vector, along with its inversion with respect to the center of its surface Brillouin zone (SBZ) for vicinal Bi surfaces.
This would be due to the strong spin-orbit interaction (SOI) of Bi.
These features suggest ideal suppression of BS, as depicted on the right side of Fig. 1(a), and thus the Q1D surface states on Bi and related compounds are regarded as a promising template to realize ballistic spin-polarized transport.

For practical application of such 1D Dirac states, tuning its electronic structure, especially to introduce the bandgap on the gapless DC, is essential for various devices such as transistors and photodetectors.
Various approaches are ongoing for this purpose, \textit{e.g.} searching for new materials with large bandgaps like transition-metal dichalcogenides \cite{Wang12}, or the use of magnetic order to open a gap in the gappless DC \cite{Xu12}.
For the electronic states of Bi, ultrathin layers grown on semiconductor substrates \cite{Hirahara06} are known to exhibit clear quantum-well states and be applicable to tune the photoelectron spin-polarization from the 2D surface states \cite{Takayama12}.

In this letter, we report the surface electronic states localized in Q1D surface of Bi fabricated on semiconductor InSb(001) substrates investigated by angle-resolved photoelectron spectroscopy (ARPES).
The surface electronic state shows a strongly anisotropic Q1D energy-contour shape with steep DC-like dispersion.
Unlike its single-crystal counterparts, a finite bandgap opening as large as 150 meV for the direct gap is observed for the Q1D states.
Moreover, spin-resolved ARPES (SARPES) reveals the spin polarization of the 1D unoccupied and occupied states, as expected for low-dimensional states with strong SOI.
The spin polarization normal to the wave vector inverts its sign in the reciprocal space with respect to the normal emission.
These results suggest that the Q1D Bi/InSb(001) surface has spin-polarized carriers with high mobility and highly efficient BS suppression, making it promising for advanced spintronic and energy-saving devices.

\begin{figure}
\includegraphics[width=80mm]{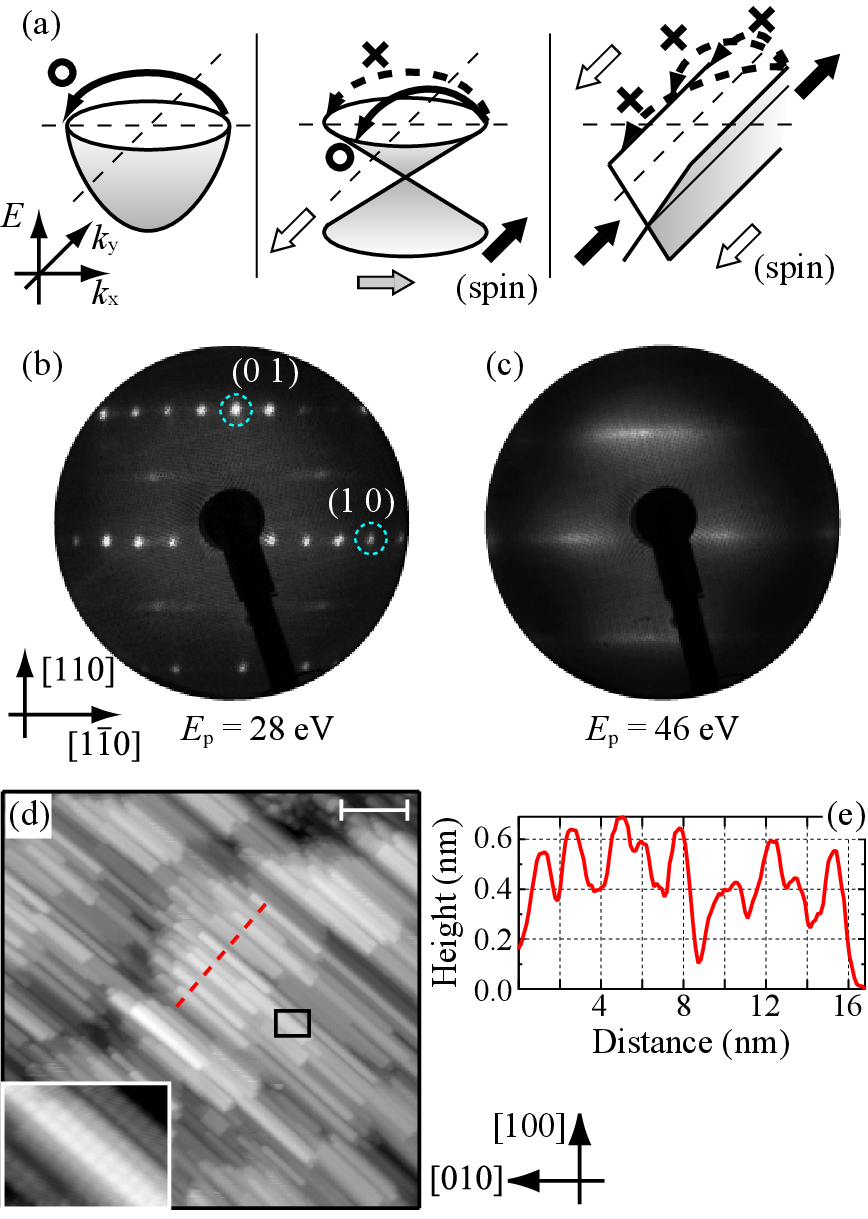}
\caption{\label{fig1}
(a) Schematics of backscattering in normal states and spin-polarized 2D/1D Dirac cones.
Low-energy electron diffraction patterns of (b) an InSb(001)-$c$(8$\times$2) substrate and (c) a Bi/InSb(001) surface covered with 10 monolayers (ML) of Bi at room temperature.
(d) Scanning tunneling microscopy (STM) image of InSb(001) surfaces covered with 5 ML of Bi at 77 K (bias voltage -1.0 V and tunneling current 0.30 nA).
The inset shows an atomically resolved image of the area in the black rectangle (bias voltage -0.10 V and tunneling current 0.15 nA).
The area of the STM image is 63$\times$63 nm (the area of the inset: 5.0$\times$3.8 nm).
The scale bar indicates 10 nm.
(e) Line profice taken along the dahsed line in (d).
}
\end{figure}

The Bi surfaces were prepared by evaporating 5 to 20 monolayers (ML) of Bi on the clean surface of InSb(001) substrates (detailed methods for sample preparation as well as measurements of surface atomic and electronic structures are provided in the supplemental materials (SM) \cite{SM}).
After Bi evaporation, the low-energy electron diffraction (LEED) pattern changed from $c$(8$\times$2) (Fig. 1(b)), consistent with the clean surface of InSb(001) \cite{Oe80}, to streaks along [1$\bar{1}$0] (Fig. 1(c)).
The inter-streak spacing along [110] is nearly the same as that between integer-order spots of the $c$(8$\times$2) pattern, suggesting that the surface lattice constant along [110] is somehow kept in the Bi layers.

Figure 1(d) shows topographic scanning tunneling microscopy (STM) images of Bi layers with thicknesses of 5 ML, showing 1D anisotropic surface structures.
Needle-like 1D structure with a width of a few atoms are fabricated on the surface.
From the atomically resolved image (the inset in Fig. 1(d)) the inter-atomic distance along [110] is estimated to be 4.5$\pm$0.1 \AA \, while that along [1$\bar{1}$0] is 4.0$\pm$0.1 \AA \, forming a rectangular unit cell.
These lattice constants are consistent with those of a Bi single crystal truncated across the bisectrix axis ((11$\bar{2}$) rhombohedral index, 4.54$\times$3.94 \AA \cite{Hofmann06}).
The line profile across the 1D structure (see Fig. 1 (e)) shows that the height of the 1D structure is $\sim$2 \AA, suggesting that these needle-like surface 1D structures are formed with corrugated single atoms as observed in vicinal surfaces of Bi single crystals \cite{Wells09, Bianchi15}.
Note that the nearest neighbour distance in the Bi single crystal is 3.1 \AA.
The difference of the Bi/InSb(001) surface from the vicinal Bi surfaces is that the spacings between the 1D corrugations show no clear periodicity along [1$\bar{1}$0].
These STM results correspond well with the LEED pattern in Fig. 1(c), which shows surface periodicity along [110] but no periodicity along [1$\bar{1}$0].
In the following part, we present the surface electronic structure of Bi films thicker than 10 ML, which is almost identical to those of the 5 ML film.
The evolution of the surface atomic and electronic structures with Bi thickness is shown in SM.

\begin{figure}
\includegraphics[width=80mm]{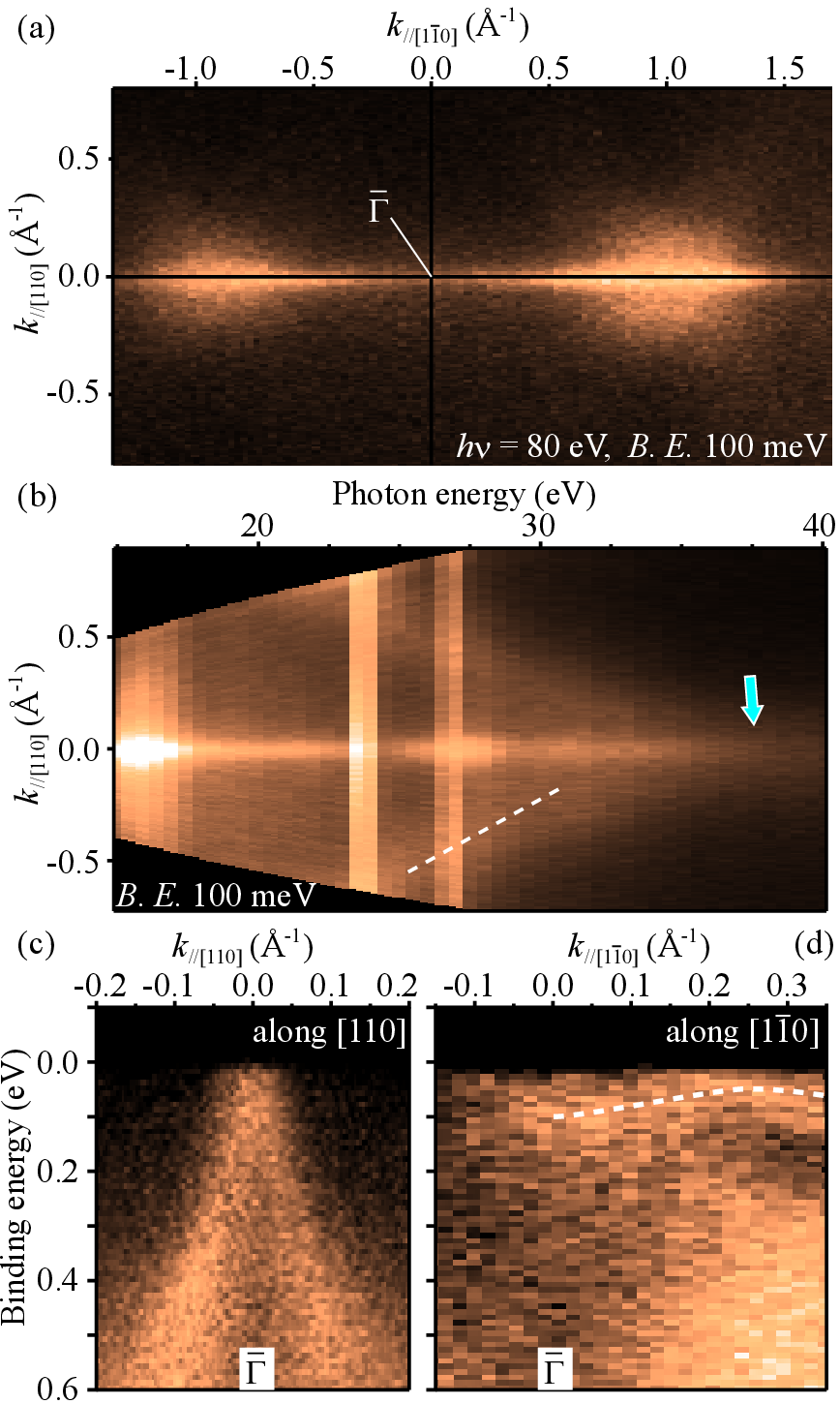}
\caption{\label{fig2}
(Color online) Electronic structure of Bi/InSb(001) measured by angle-resolved photoelectron spectroscopy (ARPES).
(a) A constant energy contour at a binding energy of 100$\pm$10 meV ($h\nu$ = 80 eV) taken at 5.5 K.
(b) Intensity plots of momentum distribution curves along [110] at $k_{//[1\bar{1}0]}$ = 0.0 \AA$^{-1}$ as a function of photon energy taken at room temperature.
ARPES intensity maps along (c) [110] and (d) [1$\bar{1}$0] measured by He I ($h\nu$ = 21.22 eV) at 31 K.
Dashed lines in (b) and (d) are visual guides.
The Bi thickness is 20 ML for (a) and 10 ML for (b-d).
}
\end{figure}

Figure 2 shows the electronic structure of the Q1D Bi surfaces obtained by ARPES measurements.
As shown in Fig. 2(a), the Bi/InSb(001) surface hosts a strongly anisotropic constant energy contour at a binding energy of 100 meV.
Figure 2(b) displays the photon energy dependence of the momentum distribution curves (MDCs) along [110] at $k_{//[1\bar{1}0]}$ = 0.0 \AA$^{-1}$.
Note that momentum-independent intensities around 24 and 27 eV originate from Bi 5$d$ resonances.
The MDC peak (indicated by an arrow) appears around $k_{//[110]}$ = 0.0 \AA$^{-1}$, independent of the photon energy, showing that this state is a surface state with no dispersion along the surface normal.
Combined with the energy contour shape in Fig. 2(a), the surface state has almost 1D character just below the Fermi level. 
The dashed line in Fig. 2(b) indicates another feature around the Fermi level with three-dimensional dispersion, which probably originates from bulk-like states localized in the Bi layers.

Figure 2(c) and (d) show the ARPES intensity maps along [110] and [1$\bar{1}$0], respectively.
Along [110], parallel to the 1D surface structure shown in Fig. 1(d), the surface state shows steep and almost linear dispersion as the lower part of DC.
In contrast, the dispersion of the surface state along [1$\bar{1}$0], the direction perpendicular to the 1D structure, is small.
The binding energy of the surface-state band at $\bar{\Gamma}$ is around 100 meV and disperses upwards to $k_{//[1\bar{1}0]}$ = 0.2 \AA$^{-1}$, where the band almost touches the Fermi level.
From $k_{//[1\bar{1}0]}$ = 0.2 to 0.3 \AA$^{-1}$, the band dispersion is almost flat. 
At first glance, such small, but non-zero dispersion may appear to contradict with the streaky LEED pattern as well as nearly random spacing between the stripes shown in the STM image (see Fig. 1).
However, the 1D structure of Bi is limited to the topmost atom, as observed by the STM line profile (Fig. 1(e)), and hence it is natural to assume a finite contribution from subsurface layers which could couple the 1D electronic states in each 1D arrays.
Such effect from subsurface states would cause the weak surface-state dispersion along [1$\bar{1}$0].

\begin{figure}
\includegraphics[width=80mm]{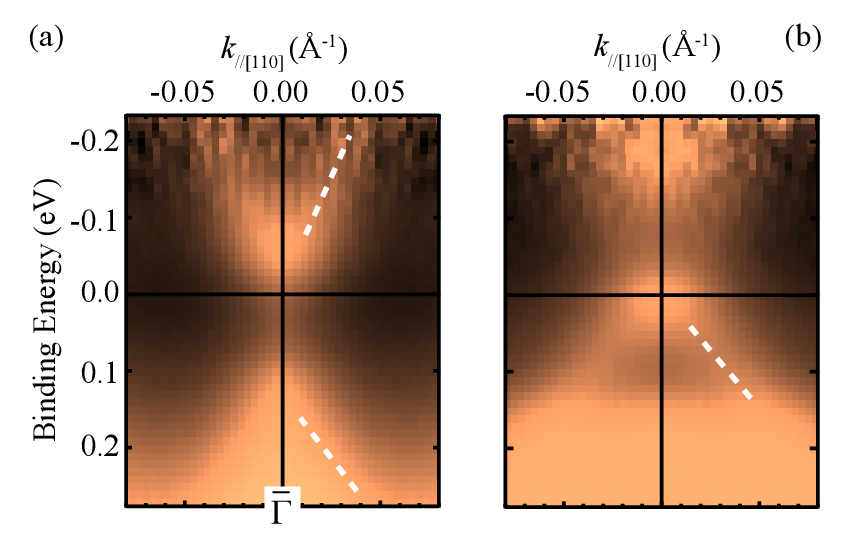}
\caption{\label{fig3}
(Color online) ARPES intensity plots of 10-ML-thick surface along [110] (h$\nu$ = 6.994 eV) at room temperature divided by the Fermi distribution function convolved with the instrumental resolution at (a) $k_{//[1\bar{1}0]}$ = 0.0 \AA$^{-1}$ and (b) 0.3 \AA$^{-1}$.
Dashed lines are visual guides.
The intensities are symmetrized with respect to $k_{//[110]}$ = 0.0 \AA$^{-1}$.
}
\end{figure}

Figure 3(a) and (b) are ARPES intensity maps at $k_{//[1\bar{1}0]}$ = 0.0 and 0.3 \AA$^{-1}$, respectively, obtained at room temperature.
To observe the surface-state dispersion above the Fermi level ($E_{\rm F}$), the ARPES spectra were divided by the Fermi distribution function convolved with the instrumental resolution.
These spectra reveal that the Q1D surface state has steep DC-like dispersion along [110] with a direct bandgap as large as $\sim$150 meV between the valence-band maximum (VBM) and the conduction-band minimum of the Q1D surface bands.
Such direct bandgap opens at least 100 meV at each $k_{//[1\bar{1}0]}$ position between  0.0 and 0.4 \AA$^{-1}$ (the detailed data is shown in SM), while the indirect gap decreases to $\sim$50 meV between 0.0 and 0.4 \AA$^{-1}$ due to the small Q1D-state dispersion along $[1\bar{1}0]$.
The nearly flat dispersion of VBM shown in Fig. 2 (d) strongly suggests that the bandgap opens at any points in this region of the reciprocal space.
Note that no other surface state dispersing across the Fermi level was observed in the other area of the reciprocal space probed in this work.
We did not find any significant change of the electronic states for the Bi/InSb(001) samples thicker than 5 ML (see SM for details).
Such behaviour is different from the thickness-dependent modification of the Bi surface electronic structure on the other substrates \cite{Takayama12}.

Although similar DC-like surface states were observed on vicinal Bi single crystals \cite{Wells09, Bianchi15}, no gap at the Dirac point has been reported.
Such gap would come from the different surface atomic structure.
The vicinal surfaces showed uniform long-range order both along and perpendicular to the 1D surface array.
In contrast, the 1D Bi arrays with its width of a few atoms on Bi/InSb(001) show no surface periodicity along $[1\bar{1}0]$ as shown in Fig. 1 (c, d) (the case is similar for the thicker films, as shown in SM).
While they are not completely isolated to each other as indicated by the small dispersion along $k_{//[1\bar{1}0]}$ (Fig. 2 (d)), such random surface corrugation could perturb the surface states forming the gap on it.
In order to consider this possibility to tune the bandgap on the Q1D states of Bi Q1D surfaces, research on the other substrates, possibly the other III-V semiconductors with similar surface atomic structure but slightly different lattice constants, is desirable.

\begin{figure}
\includegraphics[width=80mm]{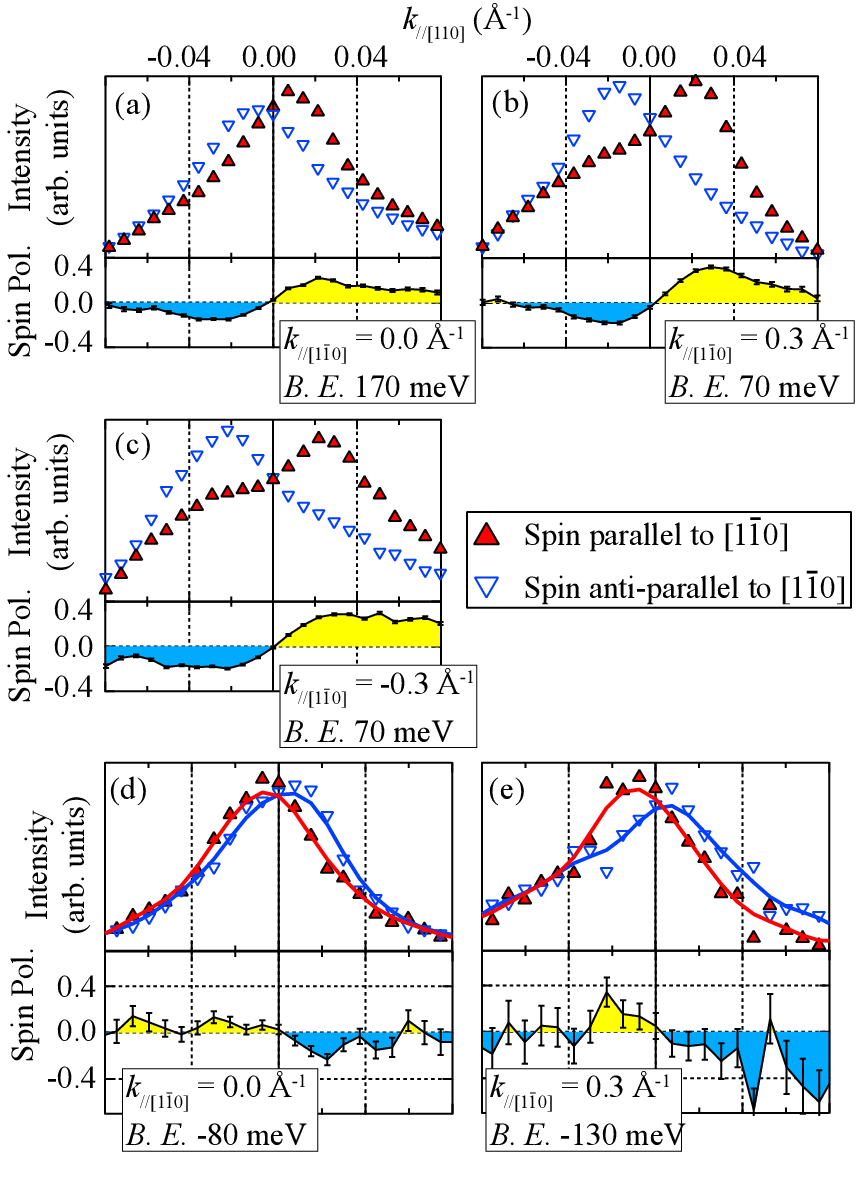}
\caption{\label{fig4}
(Color online)
Spin-resolved momentum distribution curves (MDCs) along [110] (top) and spin polarizations (bottom) at room temperature detected with a linearly polarized laser ($h\nu$ = 6.994 eV). 
The Bi thickness was 10 ML.
Binding energies and $k_{//[1\bar{1}0]}$ positions are shown.
Errors of the spin polarization values are standard statistical errors from photoelectron counting.
Solid curves in (d) and (e) are smoothed curves of spin-polarized MDCs provided as visual guides.
}
\end{figure}

To obtain further insight into the Q1D surface states, their spin polarizations were observed by SARPES at room temperature, as shown in Fig. 4.
We observed the spin-resolved MDCs along [110] at three points: $k_{//[1\bar{1}0]}$ = 0 and $\pm$0.3 \AA$^{-1}$.
The spin polarizations for each state were resolved along [1$\bar{1}$0].
As illustrated in Fig. 4(a-c), the Q1D surface states showed clear spin polarizations.
The signs of the $S_x$ spin polarization are the same at any $k_{//[1\bar{1}0]}$ and they are inverted with respect to $k_{//[110]}$ = 0.0 \AA$^{-1}$.
At room temperature, we also observed the photoelectrons from the states above $E_{\rm F}$, as shown in Fig. 3.
The spin polarizations of such states are presented in Fig. 4(d) and (e).
Although the errors for the spin polarizations of the states above $E_{\rm F}$ are much larger than those for the occupied states, finite spin polarizations are clearly observed.
The signs of the spin polarizations are the same at $k_{//[1\bar{1}0]}$ = 0 and 0.3 \AA$^{-1}$ and they are the opposite to those of the states below $E_{\rm F}$.
At $k_{//[110]}$ = 0 \AA$^{-1}$, the spin polarization of the surface band is canceled out (see SM for spin-resolved energy distribution curves).
We have also measured the spin polarization along the other orientations and found finite polarizations.
However, because of the lack of the consistency with the observed band dispersions by spin-integrated ARPES or the time-reversal symmetry, we concluded that the spin polarizations along the other directions are strongly affected by the so-called final-state effect of SARPES \cite{Jozwiak13, Kuroda16, Yaji17}.
Detailed discussion in this point is shown in SM.

These spin-polarized characteristics of the Q1D surface states make them a promising candidate for highly efficient BS suppression.
Although the BS suppression might not be perfect because the $S_x$ polarization is smaller than unity, the suppression should be better than in the 2D case depicted in Fig. 1(a).
Moreover, the Bi Q1D state should also be an ideal playground for spin-polarized carrier formation and/or tuning by photons, which is attractive for opto-spintronics.
Because the upper part of the spin-polarized surface state is unoccupied, one could excite the polarized carriers by infrared photon injection.
We hope that future quantitative research to produce and tune the spin-polarized carriers in the Q1D unoccupied states will provide further insight into opt-spintronics of low-dimensional states.

In summary, we discovered the quasi-1D electronic states on the Bi/InSb(001) surfaces with 1D needle-like surface atomic structure.
The quasi-1D states showed steep Dirac-cone-like dispersion along the 1D surface structure with a finite bandgap opening as large as 150 meV for the direct gap.
Moreover, the spin polarizations of the Q1D unoccupied and occupied states were observed. 
The spin orientation was inverted depending on the wave vector direction parallel to the 1D structure and some spin components changed significantly depending on the incident photon geometry.
These results suggest that spin-polarized carrier formation, which can possibly be tuned by external fields such as photon polarizations, and highly efficient BS suppression occurred in the Q1D surface states of Bi, making them attractive for use in future spintronic and energy-saving devices.

We acknowledge D. Ragonnet, F. Deschamps and K. Hagiwara for their support during the experiments on the CASSIOP\'EE beamline at synchrotron SOLEIL and R. Yori during the STM measurements at HiSOR.
Part of the ARPES experiments were performed under the Nanotechnology Platform Program at IMS of the Ministry of Education, Culture, Sports, Science and Technology (MEXT), Japan and HiSOR Proposal No. 16AU010.
SARPES experiment in this work was carried out by the joint research in ISSP, the University of Tokyo
This work was also supported by the JSPS Grant-in-Aid (B) (Grant No. JP26887024) and the Murata Science Foundation.

\newpage

\renewcommand{\thefigure}{\arabic{figure}S}
\setcounter{figure}{0}

\section*{Supplementary material for: Spin-polarized quasi 1D state with finite bandgap on the Bi/InSb(001) surface}

\subsection{Experimental methods}

InSb(001) substrates were cleaned by repeated cycles of sputtering and annealing up to 680 K until the $c$(8$\times$2) low-energy electron diffraction (LEED) pattern was observed.
Then, Bi was evaporated at room temperature from a Knudsen cell.
The Bi evaporation rate was estimated by a quartz microbalance as well as known LEED patterns on Si(111) \cite{Nagao04} and Ge(111) \cite{Hatta09}: In this work, one monolayer (ML) is defined as the atom density of bulk-truncated InSb(001).
Surface atomic structures before and after the Bi evaporation were observed by LEED and scanning tunneling microscopy (STM).
The STM measurement was performed by using an LT-STM (Omicron NanoTechnorogy GmbH) operated at 77 K in an ultrahigh vacuum with constant-current mode.

\begin{figure}
\includegraphics[width=80mm]{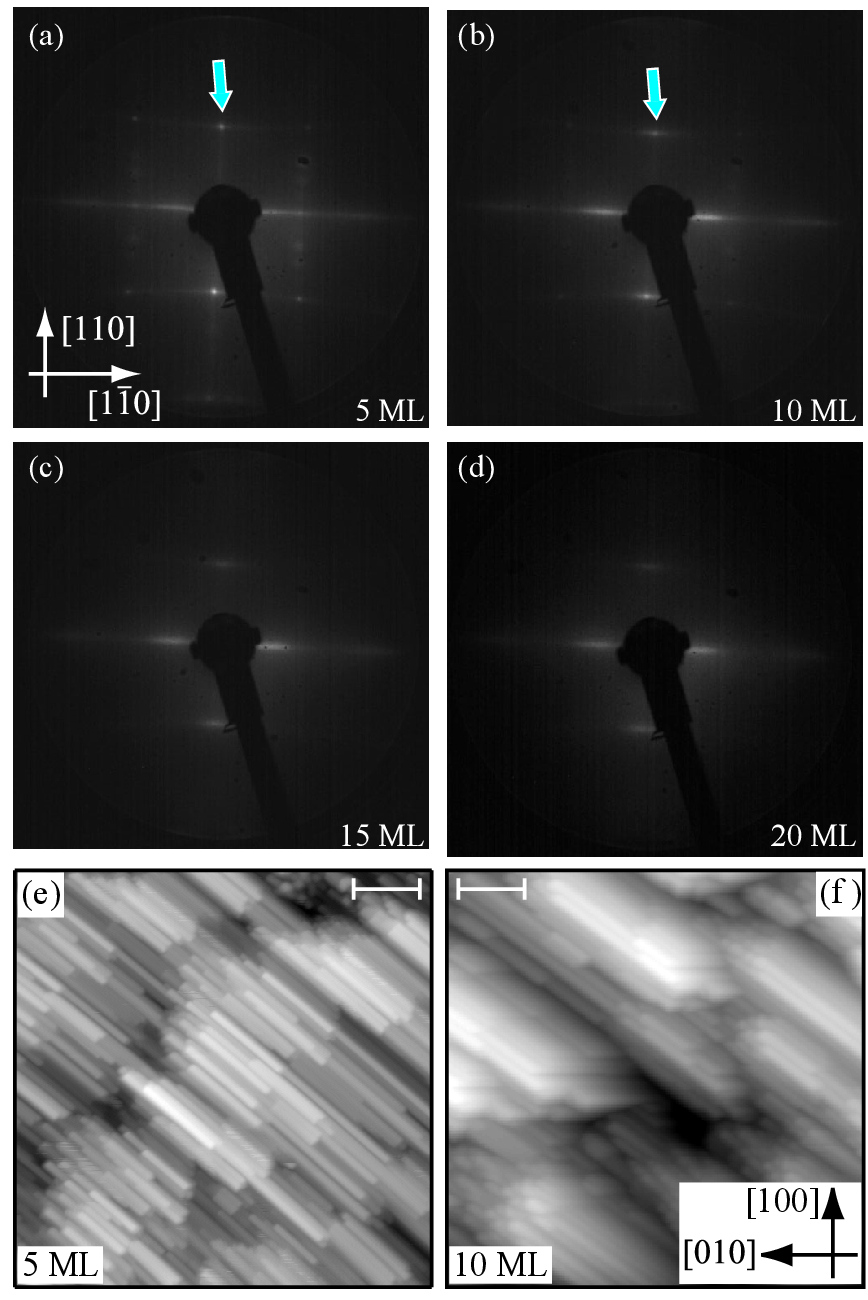}
\caption{\label{fig1s}
(a-d) Low-energy electron diffraction (LEED) patterns of the Bi/InSb(001) surfaces taken at 31 K ($E_p$ = 92 eV).
(e, f) Scanning tunneling microscopy (STM) images of the Bi/InSb(001) surfaces with the Bi coverages at (e) 5 ML (the same data as Fig. 1 (d) in the main text) and (f) 10 ML (bias voltage +1.0 V) taken at 77 K. The area of the STM images are 63$\times$63 nm$^2$ (slace bars: 10 nm).
}
\end{figure}

The surface electronic structures of the Bi/InSb(001) surfaces and their spin polarizations were measured with angle-resolved photoelectron spectroscopy (ARPES) and spin-resolved ARPES (SARPES).
ARPES measurements were performed with a He lamp, synchrotron radiation at the CASSIOP\'EE beamline of Synchrotron SOLEIL (photon energies ranged from 15 to 80 eV), and a laser system providing 6.994-eV photons at the Institute for Solid-State Physics (ISSP), the University of Tokyo.
The overall energy resolutions were better than 10 meV with the He lamp and laser and 20 meV for synchrotron radiation.
The SARPES measurements were performed using the laser-SARPES machine at ISSP \cite{Yaji16}.
The energy resolution of the laser-SARPES setup was better than 20 meV and the Sherman function was set to 0.3.


\subsection{Atomic and electronic structure depending on the amout of evaporated Bi}

Figure S1 is the LEED patterns and STM images of the Bi nanoribbons grown on InSb(001) substrates at various thicknesses.
As shown in the LEED patterns, the streaks along [1$\bar{1}$0] are observed at any thicknesses.
At 5 ML, there are the other weak features along [110].
These features would be due to the co-existing region with thinner Bi thickness: possibly (1$\times$3) or $c$(2$\times$6) surfaces \cite{Laukkanen10, Ohtsubo15} with Bi thickness around 1 ML.
Since the Bi/InSb(001) surfaces studied in this work were not annealed to high temperature, it is natural that the surface is not completely homogenious with variety of the Bi thickness, as observed in the scanning tunneling microscopy (STM) images in Fig. S1 (e) and 1 (f).
At 5 and 10 ML, one can also find sharp spots as indicated by the arrows in Figs. S1 (a) and S1 (b).
These spots appears at the same positions as the integer-order spots of the InSb(001) substrates.
They would come from InSb(001) substrates, since they disappear at higher thicknesses.

\begin{figure}
\includegraphics[width=75mm]{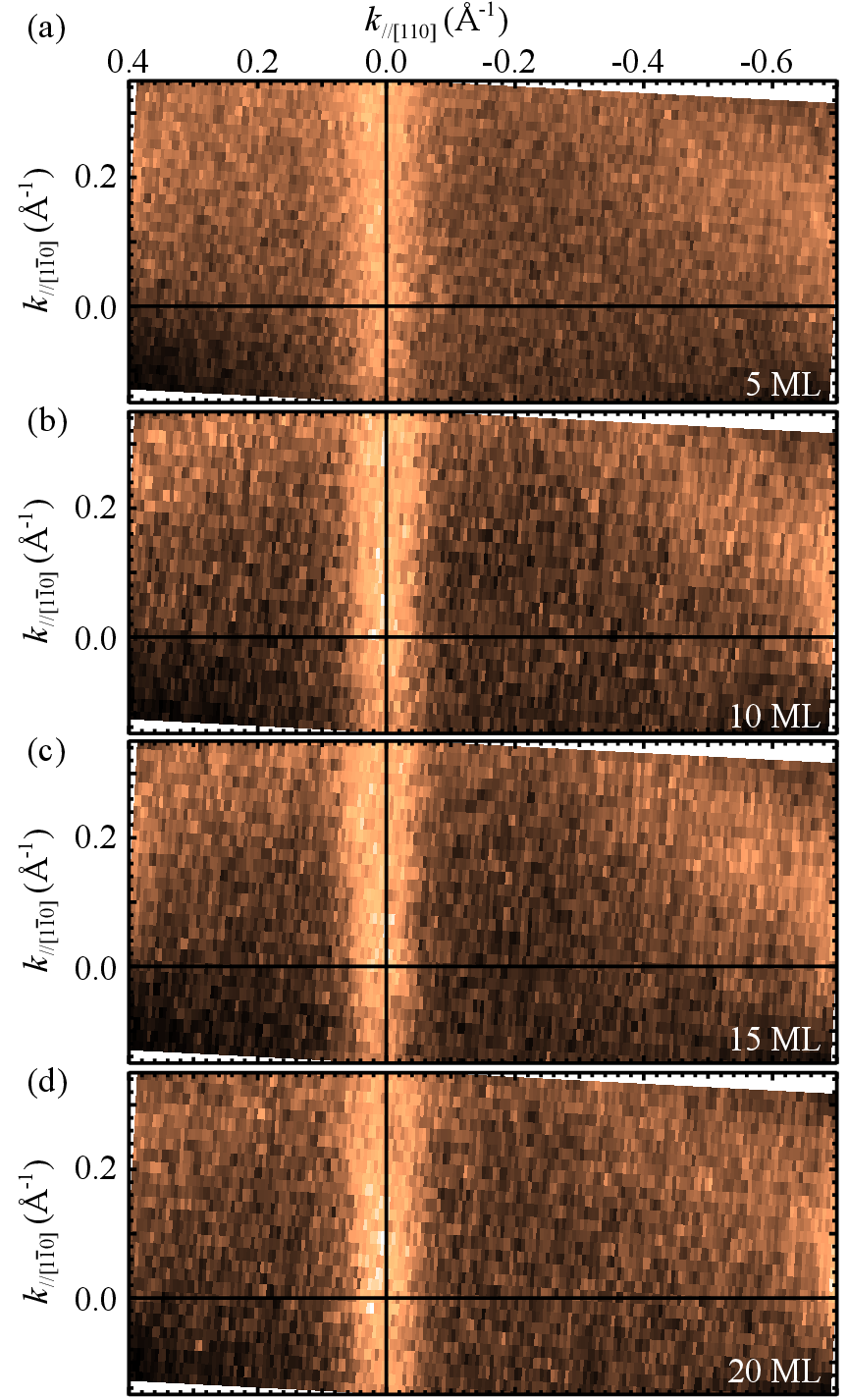}
\caption{\label{fig2s}
ARPES constant energy contours at the binding energy of 100$\pm$10 meV ($h\nu$ = 21.22 eV) taken at 31 K.
}
\end{figure}

Figures S1 (e) and (f) shows the STM images at 5 and 10 ML.
The blurred image at 10 ML is caused by the different STM-tip condition.
At 10 ML, the 1D surface array tends to bunch together forming larger corrugations on the surface than those on the 5 ML surface.
However, one can find the common feature at both surfaces, the 1D surface structure with its width of a few atoms.
As shown in the following, the electronic structure showed no significant change depending on the Bi thickness above 5 ML.
It suggests that this 1D array is the essential surface atomic structure for the quasi-1D Dirac-like surface states discussed in this work.

\begin{figure}
\includegraphics[width=80mm]{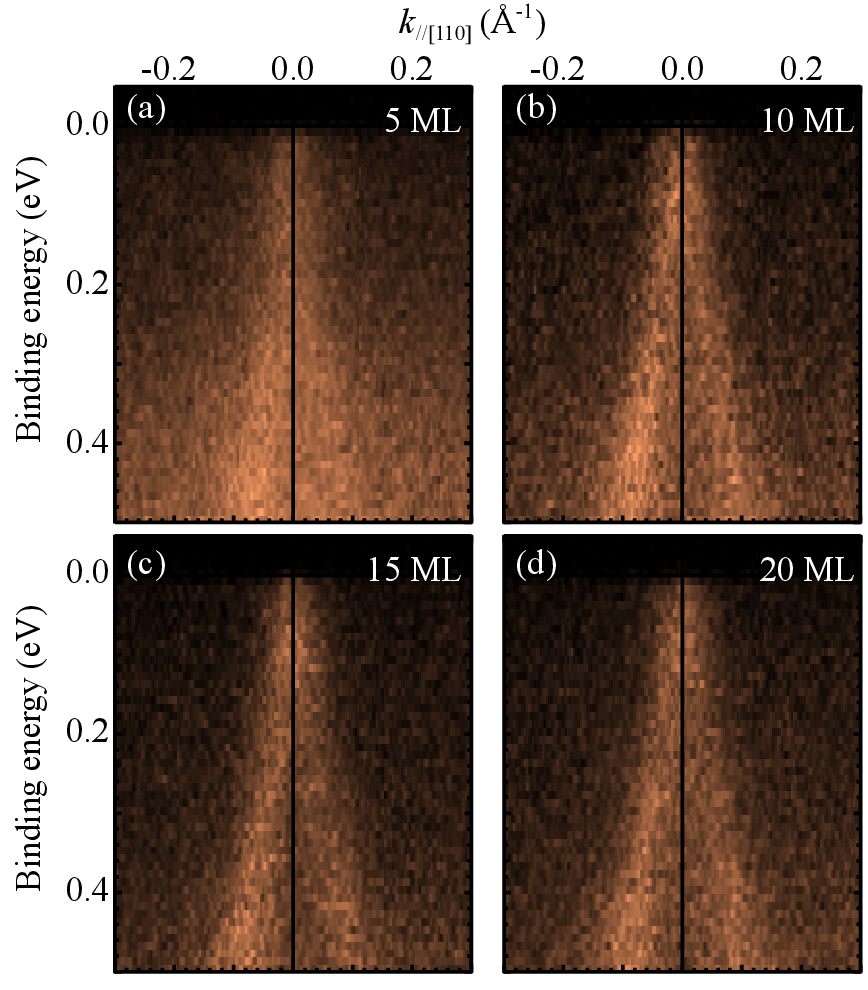}
\caption{\label{fig3s}
ARPES intensity plots along [110] at $k_{//[1\bar{1}0]}$ = 0.0 \AA$^{-1}$ taken at 31 K ($h\nu$ = 21.22 eV).
}
\end{figure}

Figure S2 shows the constant energy contours around the binding energy of 100 meV by angle-resolved photoelectron spectroscopy (ARPES) with 5-20 ML Bi.
As shown there, the one-dimensional (1D) feature at $k_{[110]}$ = 0 \AA$^{-1}$ does not change depending on the evaporated Bi amounts.
The 1D feature at 5 ML is a little weaker than the others, possibly because of the coexistence with the other surfaces as observed by LEED (Fig. S1 (a)).
However, from ARPES measurements, no features derived from the other surfaces, such as metallic states on Bi/InSb(001)-(1$\times$3) \cite{Ohtsubo15}, were observed around the Fermi level.
Therefore, the most area of the surface should be covered with Bi nanoribbons even at 5 ML.
Above 10 ML, one cannot find any difference between each other from the constant energy contours.
In addition to the 1D surface state, there is a broad intensity around ($k_{[110]}$, $k_{[1\bar{1}0]}$) = (-0.6 \AA$^{-1}$, 0.2 \AA$^{-1}$) at any thicknesses.
Such feature was not observed at the other photon energies (see Fig. 2 (a) in the main text) and hence it would be due to the three-dimensional, bulk-like states of the Bi layers.

Figure S3 is the ARPES intensity plots along [110] at $k_{//[1\bar{1}0]}$ = 0.0 \AA$^{-1}$ with 5-20 ML Bi.
In general, the 1D surface electronic structure with Dirac-cone-like steep dispersion is the same at any Bi thicknesses, consistent with what are observed from constant energy contours in Fig. S2.
The dispersion at 5 ML is more diffuse than the others and it would be due to the limtied area of the Bi 1D structure at this thickness.
The thicker samples show almost identical dispersions.

In summary, all the results from LEED, STM, and ARPES suggest that the 1D surface state with the Bi thickness above 10 ML is not influenced by the thickness of the Bi layers.
The surface atomic/electronic structures of 5 ML is almost identical to but slightly weaker than those at higher thicknesses.
It suggests that the Q1D Bi/InSb(001) structure requires the Bi average thickness above 10 ML to cover the whole surface.

\begin{figure}
\includegraphics[width=80mm]{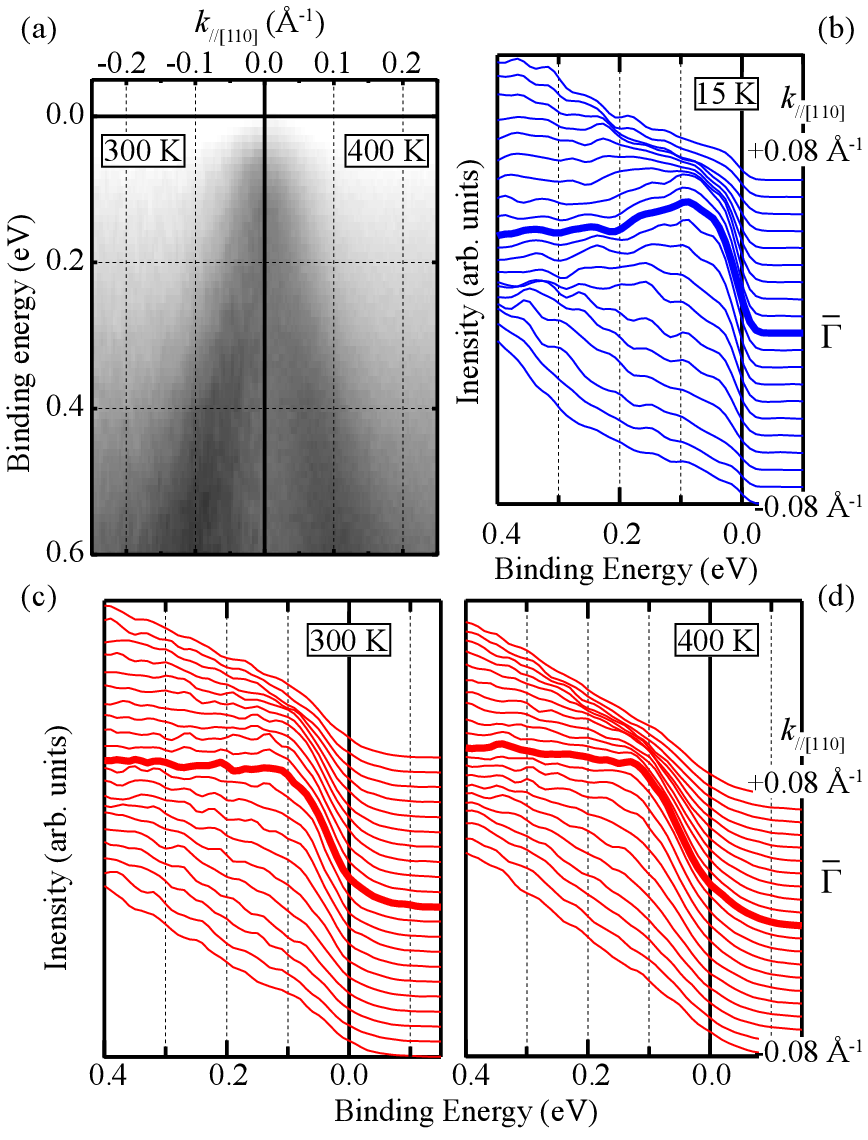}
\caption{\label{fig4s}
ARPES data taken with He lamp (h$\nu$ = 21.2 eV). (a) An ARPES intensity plot at 300 K (left) and 400 K (right) at $k_{//[1\bar{1}0]}$ = 0.0 \AA$^{-1}$. (b-d) ARPES energy distribution curves (EDCs) measured (b) at 15 K, (c) 300 K, and (d) 400 K.
}
\end{figure}

\subsection{Temperature dependence of surface state}

Figure S4 (a) shows the 2D ARPES intensity plot taken at 300 and 400 K at $k_{//[1\bar{1}0]}$ = 0.0 \AA$^{-1}$.
As shown there, the dispersion of the Q1D state is almost the same at higher temperature, comparing with those at low temperature (31 K or lowewr, in this work).
To be more precise, we compared the ARPES energy distribution curves (EDCs) at each temperature (Figs. S4 (b-d)).
As shown there, the position of the valence-band maximum (VBM) at $k_{//[110]}$ = 0 \AA$^{-1}$ slightly moves downwards ($\sim$50 meV) at high temperature.
Note that the VBM at 15 K is still away from the Fermi level and such a small energy shift does not make any changes in the whole discussion in the main text; Dirac-cone-like steep and quasi-1D surface bands with finite bandgap and spin polarization.


\subsection{Finite bandgap on the Q1D states along [1$\bar{1}$0]}

Figure S5 (a) and (b) show the ARPES EDCs along [110] at $k_{//[1\bar{1}0]}$ = 0.0 \AA$^{-1}$.
They indicate the finite bandgap at 300-400 K.
While the signal from the upper cone is weak and noisy with He lamp ((h$\nu$ = 21.2 eV, Fig. S5 (b)), one can observe the similar dispersion as observed by laser (h$\nu$ = 6.994 eV, Fig. S5 (a)), suggesting that the upper cone is also a surface state without dispersion along the surface normal.
This different intensity would be due to the photoexcitation coross section depending on the incident photon energies.

Figure S5 (c) is the energy distribution curves integrated in $k_{//[110]}$ = 0.00$\pm$0.02 \AA$^{-1}$ from the intensity plots divided with the Fermi distribution function.
While the signal from the upper cone is not so strong as those obtained by laser (Fig. 3 in the main text), one can clearly see the unoccupied state around -0.1 eV at  $k_{//[1\bar{1}0]}$ = 0.0 and 0.1 \AA$^{-1}$.
In the larger $k_{//[1\bar{1}0]}$ region, the peak position of the unoccupied surface state is not clear because of the limited thermal population of the electronic states above the Fermi level, even at 400 K.
However, it is at least clear that the gap between the valence-band maximum and conduction-band minimum of the Q1D suface bands does not close in this region.
From this data, it is proved that the Q1D Dirac-like state studied in this work holds finite bandgap in the wide range of the reciplocal space.

\begin{figure}
\includegraphics[width=80mm]{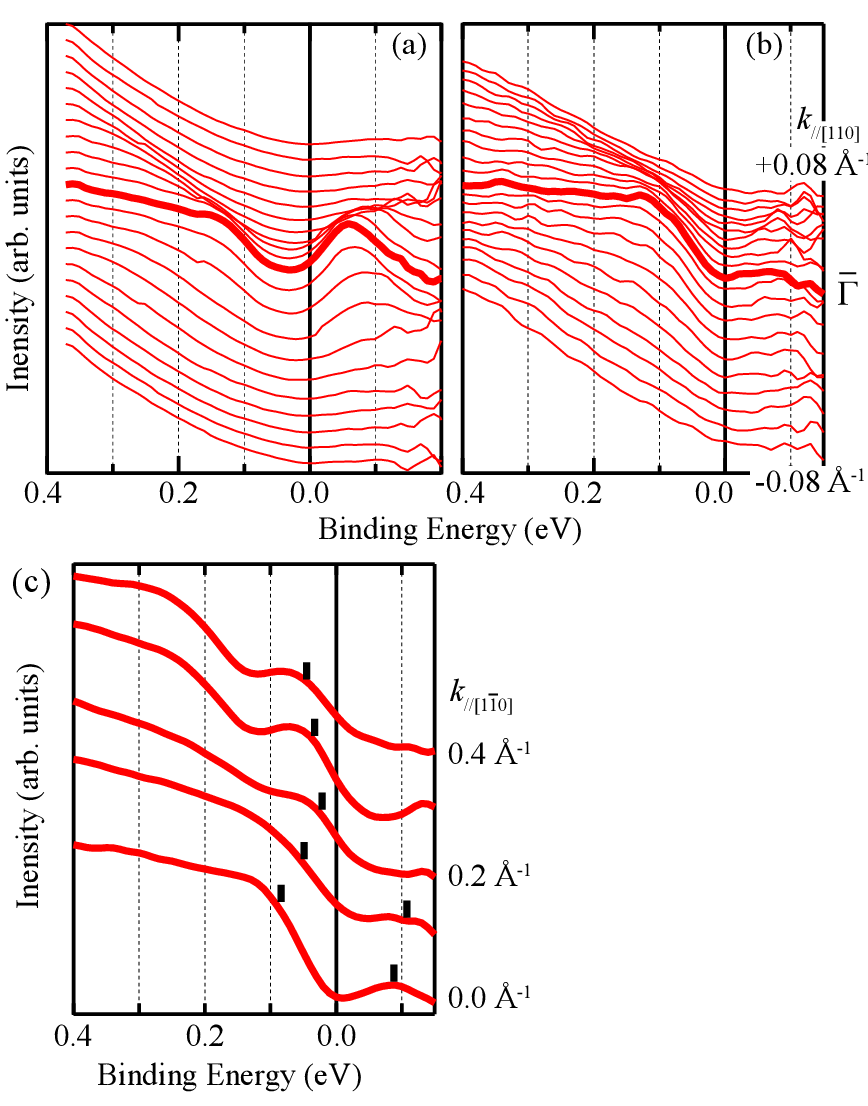}
\caption{\label{figSN5}
(a) ARPES EDCs takan along [110] at $k_{//[1\bar{1}0]}$ = 0.0 \AA$^{-1}$ (h$\nu$ = 6.994 eV) at room temperature. Each spectrum is divided by the Fermi distribution function convolved with the instrumental resolution.
(b) The same as (a) but at 400 K with He lamp (h$\nu$ = 21.2 eV).
(c) The same as (b) but takan along [1$\bar{1}$0]. Each spectrum is intagrated within $k_{//[110]}$ = 0.00$\pm$0.02 \AA$^{-1}$. Ticks indicate the peak positions of the quasi-1D surface bands.
}
\end{figure}

\subsection{Spin-resolved energy distribution curves}

Figure S6 (a) shows the spin-resolved ARPES energy distribution curves (EDCs) along [110].
The spin polarizations along [1$\bar{1}$0] were resolved here.
The Dirac-cone-like dispersion of the surface state is shown in Fig. S{6} (b), which is constructed by the average of the spin-resolved EDCs shown in Fig. S{6} (a).
Away from $k_{//[1\bar{1}0]}$ = 0.0 \AA$^{-1}$, one can find clear spin polarizations along [110] whose energy positions are consistent with the surface bands shown in Fig. S{6} (b).
The spin polarizations of the surface bands inverts with respect to $k_{//[1\bar{1}0]}$ = 0.0 \AA$^{-1}$.
These behaviors agree well with the spin-resolved MDCs in the main text (Figure 4).
At the top of the surface band, where the surface state almost touches the Fermi level, the spin polarization disappears.

\begin{figure}[t]
\includegraphics[width=80mm]{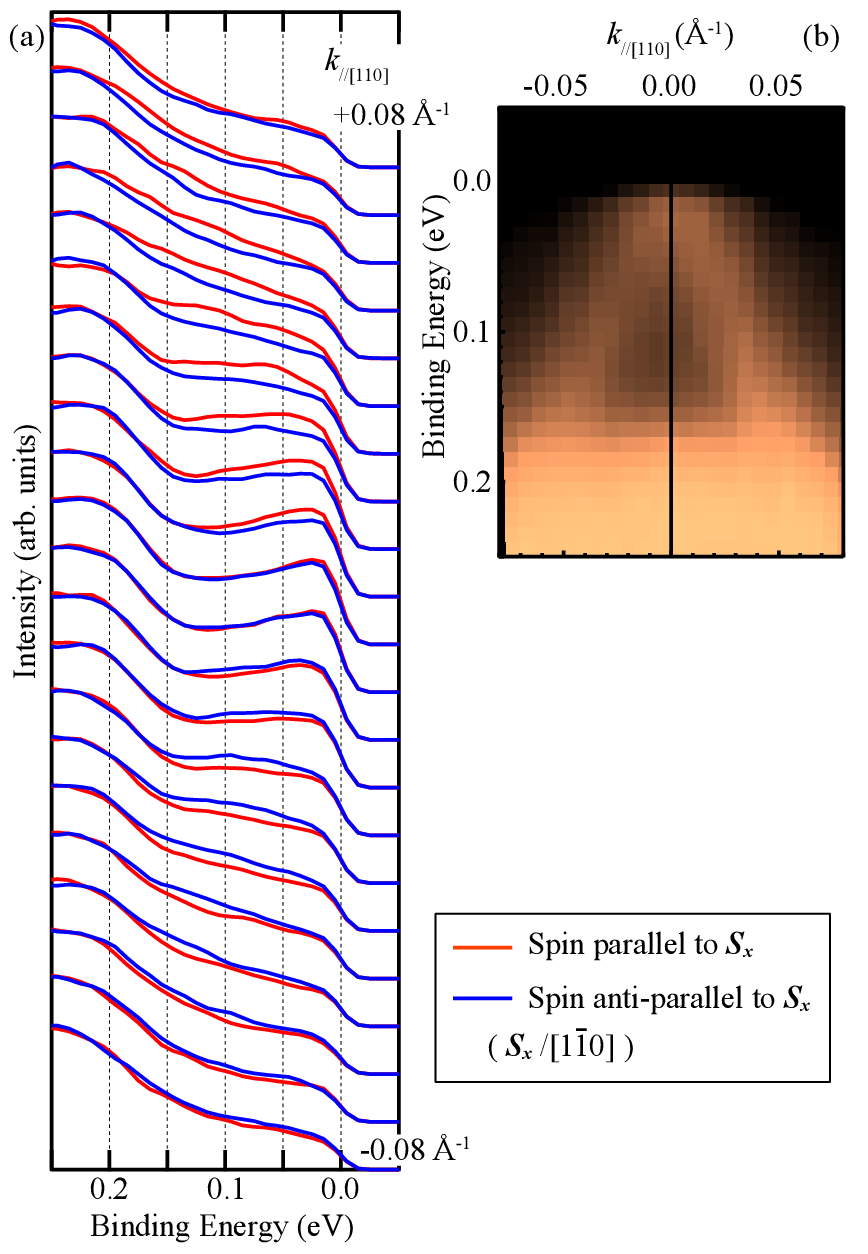}
\caption{\label{fig6s}
(a) Spin-resolved ARPES energy distribution curves taken along [110] at 15 K with linearly polarized laser (h$\nu$ = 6.994 eV). $k_{//[1\bar{1}0]}$ = 0.25 \AA$^{-1}$.
(b) The spin-integrated ARPES intensity plot constructed by the EDCs shown in (a).
}
\end{figure}

\begin{figure*}[t]
\includegraphics[width=150mm]{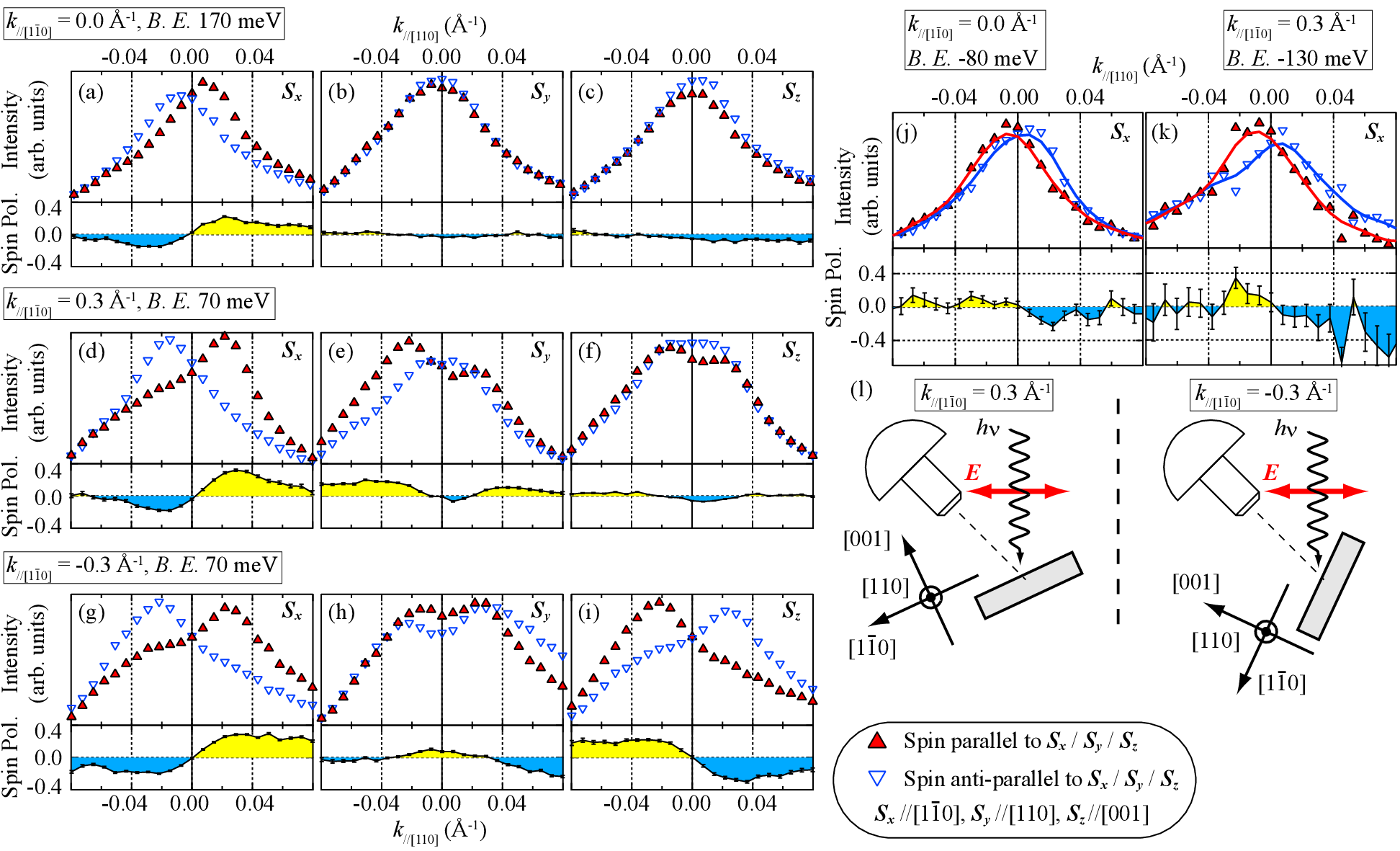}
\caption{\label{fig7s}
(a-k) Spin-resolved momentum distribution curves (MDCs) along [110] (top) and spin polarizations (bottom) at room temperature detected with a linearly polarized laser ($h\nu$ = 6.994 eV). ((a), (d), (g), (j) and (k) are the same as those shown in Fig. 4 in the main text).
The Bi thickness was 10 ML.
Binding energies and $k_{//[1\bar{1}0]}$ positions are shown.
Errors of the spin polarization values are standard statistical errors from photoelectron counting.
Solid curves in (j) and (k) are smoothed curves of spin-polarized MDCs provided as visual guides.
(l) Schematics of SARPES experimental geometries.
}
\end{figure*}

\subsection{Spin polarization of photoelectrons along various orientations}

Figure S{7} shows the spin-resolved MDCs along [110] at three points: $k_{//[1\bar{1}0]}$ = 0 and $\pm$0.3 \AA$^{-1}$.
The spin polarizations for each state were resolved along $S_x$ (//[1$\bar{1}$0]), $S_y$ (//[110]), and $S_z$ (//[001]).
The MDCs along $S_x$ are the same as those shown in the main text (Fig. 4).

Along the $S_y$ and $S_z$ orientations, the polarizations are almost negligible at $k_{//[1\bar{1}0]}$ = 0.0 \AA$^{-1}$ but become evident at $k_{//[1\bar{1}0]}$ = $\pm$0.3 \AA$^{-1}$.
The spin polarization along $S_y$ does not correspond to the surface band dispersion, as shown in Fig. 3(b) and the MDC peak positions in Fig. S{\red 7} (d) and (g).
Meanwhile, spin polarization towards $S_z$ becomes prominent at $k_{//[1\bar{1}0]}$ = -0.3 \AA$^{-1}$, while it is almost negligible at the opposite side of the SBZ, $k_{//[1\bar{1}0]}$ = +0.3 \AA$^{-1}$.
These unusual spin textures along $S_y$ and $S_z$ cannot be obtained solely from the spin polarization of the initial states.
These spin polarizations would originate from the photoexcitation process, the so-called final-state effect.
It is known that the spin polarization of photoelectrons strongly depends on the polarization of incident photons as well as the photon energies, as observed for various spin-polarized surface states \cite{Jozwiak13, Kuroda16, Yaji17}.
As depicted in Fig. S{7}(l), we changed the polar angle of the substrate for different $k_{//[1\bar{1}0]}$ points, causing the in-plane (out-of-plane) component of the electric field from incident photons to become dominant for positive (negative) $k_{//[1\bar{1}0]}$ values.
The drastic change of the $S_z$ spin component at $k_{//[1\bar{1}0]}$ = $\pm$0.3 \AA$^{-1}$ might be related to these different incident electric fields.
Based on this framework, one may think that the spin polarizations along $S_x$ are also an artifact due to the final-state effect.
However, the $S_x$ spin polarization values does not change so much at any $k_{//[1\bar{1}0]}$ (that is, with different photon-incident geometries), and thus these $S_x$ spin polarizations would originates from the initial states.
Actually, the spin polarizations along $S_x$ were observed from the similar but gapless Q1D Dirac-cone-like states on Bi vicinal surfaces measured with different experimental geometry as well as the photon energies \cite{Wells09, Bianchi15}.
It also supports the origin of the spin polarization along $S_x$ from its initial state.

As shown in the $S_z$ polarization in Fig. S{\red 7}(f) and (i), the spin polarization orientations of the Q1D surface states are strongly influenced by the incident photon polarization.
It was recently shown that the spin polarization of photoelectrons can be controlled by incident photon polarizations with large degree of freedom \cite{Yaji17} and it would be natural to assume that the same effect is applicable for photo-induced carrier formation by infrared photons.
Thanks to this, one might be able to tune the spin polarization of photoexcited electrons by controlling the polarization of pump (infrared) photons.


\begin{thebibliography}{99}
\bibitem{Geim07}
A. K. Geim and K. S. Novoselov, Nature Mat. {\bf 6}, 183 (2007)
\bibitem{Hasan10}
M. Z. Hasan and C. L. Kane, Rev. Mod. Phys. {\bf 82}, 3045 (2010).
\bibitem{Manchon15}
A. Manchon \textit{et al.}, Nature Mat. {\bf 14}, 871 (2015).
\bibitem{Roushan09}
Roushan \textit{et al.}, Nature {\bf 460}, 1106 (2009).
\bibitem{Kim14}
S. Kim \textit{et al.}, Phys. Rev. Lett. {\bf 112}, 136802 (2014).
\bibitem{Konig07}
M. K\"{o}nig \textit{et al.}, Science {\bf 318}, 766 (2007).

\bibitem{Wells09}
J. W. Wells \textit{et al.}, Phys. Rev. Lett. {\bf 102}, 096802 (2009).
\bibitem{Bianchi15}
M. Bianchi \textit{et al.}, Phys. Rev. B {\bf 91}, 165307 (2015).
\bibitem{Autes16}
G. Aut\`{e}s \textit{et al.}, Nature Mat. {\bf 15}, 154 (2016).

\bibitem{Wang12}
Q.-H. Wang \textit{et al.}, Nature Nanotech. {\bf 7}, 699 (2012).
\bibitem{Xu12}
S.-Y. Xu \textit{et al.}, Nature Mat. {\bf 8}, 616 (2012).

\bibitem{Hirahara06}
T. Hirahara \textit{et al.}, Phys. Rev. Lett. {\bf 97}, 146803 (2006).
\bibitem{Takayama12}
A. Takayama \textit{et al.}, Nano Lett. {\bf 12}, 1776 (2012).

\bibitem{SM}
See Supplemental Material for the detailed sample preparation procedure and additional ARPES/spin-resolved ARPES dataset, which includes Refs. \cite{Nagao04, Hatta09, Laukkanen10, Ohtsubo15, Yaji16}.

\bibitem{Nagao04}
T. Nagao \textit{et al.}, Phys. Rev. Lett. {\bf 93}, 105501 (2004).
\bibitem{Hatta09}
S. Hatta \textit{et al.}, Appl. Surf. Sci. {\bf 256}, 1252 (2009).
\bibitem{Laukkanen10}
P. Laukkanen \textit{et al.}, Phys. Rev. B {\bf 81}, 035310 (2010).
\bibitem{Ohtsubo15}
Y. Ohtsubo \textit{et al.}, Phys. Rev. Lett. {\bf 115}, 256404 (2015).
\bibitem{Yaji16}
K. Yaji \textit{et al.}, Rev. Sci. Instrum. {\bf 87}, 053111 (2016).

\bibitem{Oe80}
K. Oe, S. Ando and K. Sugiyama, Jpn. J. Appl. Phys. {\bf 19}, L417 (1980).


\bibitem{Hofmann06}
Ph. Hofmann, Prog. Surf. Sci. {\bf 81}, 191 (2006).

\bibitem{Jozwiak13}
C. Jozwiak \textit{et al.}, Nat. Phys. {\bf 9}, 293 (2013).
\bibitem{Kuroda16}
K. Kuroda \textit{et al.}, Phys. Rev. B {\bf 94}, 165162 (2016).
\bibitem{Yaji17}
K. Yaji \textit{et al.}, Nature Commun. {\bf 8}, 14588 (2017).
\end{thebibliography}

\begin{thebibliography}{99}
\bibitem{Nagao04}
T. Nagao \textit{et al.}, Phys. Rev. Lett. {\bf 93}, 105501 (2004).
\bibitem{Hatta09}
S. Hatta \textit{et al.}, Appl. Surf. Sci. {\bf 256}, 1252 (2009).
\bibitem{Yaji16}
K. Yaji \textit{et al.}, Rev. Sci. Instrum. {\bf 87}, 053111 (2016).
\bibitem{Laukkanen10}
P. Laukkanen \textit{et al.}, Phys. Rev. B {\bf 81}, 035310 (2010).
\bibitem{Ohtsubo15}
Y. Ohtsubo \textit{et al.}, Phys. Rev. Lett. {\bf 115}, 256404 (2015).

\bibitem{Jozwiak13}
C. Jozwiak \textit{et al.}, Nat. Phys. {\bf 9}, 293 (2013).
\bibitem{Kuroda16}
K. Kuroda \textit{et al.}, Phys. Rev. B {\bf 94}, 165162 (2016).
\bibitem{Yaji17}
K. Yaji \textit{et al.}, Nature Commun. {\bf 8}, 14588 (2017).

\bibitem{Wells09}
J. W. Wells \textit{et al.}, Phys. Rev. Lett. {\bf 102}, 096802 (2009).
\bibitem{Bianchi15}
M. Bianchi \textit{et al.}, Phys. Rev. B {\bf 91}, 165307 (2015).

\end{thebibliography}
\end{document}